\documentclass[aps,prd,showpacs,superscriptaddress,preprint,raggedfooter]{revtex4-1}
\usepackage[utf8]{inputenc}

\usepackage{dcolumn}   % needed for some tables
\usepackage{tabulary}  % for customizing width of the tables
\usepackage{array}     % for tables
\newcolumntype{K}[1]{>{\centering\arraybackslash}p{#1}} % for tables
\usepackage{bm}        % for math
\usepackage{amssymb}   % for math
\usepackage{amsmath}   % for math
\usepackage{mathrsfs}  % some math fonts

\usepackage{float}
\usepackage{graphicx}
\usepackage{epstopdf}
\usepackage{subfigure}
\usepackage{hyperref}
\usepackage{color}
\usepackage{epsfig}

\begin{document}

\title{Collision of $\phi^4$ kinks free of the Peierls--Nabarro barrier \\ in the regime of strong discreteness}

\author{Alidad Askari}
\email{alidadaskari@gmail.com}
\affiliation{Department of Physics, Faculty of Science, University of Hormozgan, P.O.Box 3995, Bandar Abbas, Iran}

\author{Aliakbar Moradi Marjaneh}
\email{moradimarjaneh@gmail.com}
\affiliation{Young Researchers and Elite Club, Quchan Branch, Islamic Azad University, Quchan, Iran}

\author{Zhanna~G.~Rakhmatullina}
\email{rakhzha@gmail.com}
\affiliation{Institute for Metals Superplasticity Problems, Russian Academy of Sciences, Ufa 450001, Russia}

\author{Mahdy Ebrahimi-Loushab}
\email{ebrahimi.mahdy@gmail.com}
\affiliation{Department of Physics, Faculty of Montazeri Technical and Vocational University (TVU), Khorasan Razavi, IRAN}

\author{Danial Saadatmand}
\email{saadatmand.d@gmail.com}
\affiliation{Department of Physics, University of Sistan and Baluchestan, Zahedan, Iran}

\author{Vakhid~A.~Gani}
\email{vagani@mephi.ru}
\affiliation{Department of Mathematics, National Research Nuclear University MEPhI (Moscow Engineering Physics Institute), Moscow 115409, Russia}
\affiliation{Theory Department, Institute for Theoretical and Experimental Physics of National Research Centre ``Kurchatov Institute'', Moscow 117218, Russia}

\author{Panayotis~G.~Kevrekidis}
\email{kevrekid@umass.edu}
\affiliation{ Department of Mathematics and Statistics, University of Massachusetts, Amherst, Massachusetts 01003, USA}
\affiliation{Mathematical Institute, University of Oxford, OX26GG, UK}

\author{Sergey~V.~Dmitriev}
\email{dmitriev.sergey.v@gmail.com}
\affiliation{Institute of Molecule and Crystal Physics, Ufa Federal Research Center of Russian Academy of Sciences, Ufa 450075, Russia}
\affiliation{Institute of Mathematics with Computing Centre, Ufa Federal Research Centre of Russian Academy of Sciences, Ufa 450000, Russia}

\begin{abstract}
The two major effects observed in collisions of the {\em continuum} $\phi^4$ kinks are (i)~the existence of critical collision velocity above which the kinks always emerge from the collision and (ii)~the existence of the escape windows for multi-bounce collisions with the velocity below the critical one, associated with the energy exchange between the kink's internal and translational modes. The potential merger (for sufficiently low collision speeds) of the kink and antikink produces a bion with oscillation frequency $\omega_{\rm B}^{}$, which constantly radiates energy, since its higher harmonics are always within the phonon spectrum. Similar effects have been observed in the discrete $\phi^4$ kink-antikink collisions for relatively weak discreteness. Here we analyze kinks colliding with their mirror image antikinks in the regime of strong discreteness considering an exceptional discretization of the $\phi^4$ field equation where the static Peierls--Nabarro potential is precisely zero and the not-too-fast kinks can propagate practically radiating no energy. Several new effects are observed in this case, originating from the fact that the phonon band width is small for strongly discrete lattices and for even higher discreteness an inversion of the phonon spectrum takes place with the short waves becoming low-frequency waves. When the phonon band is narrow, not a bion but a discrete breather with frequency $\omega_{\rm DB}^{}$ and all higher harmonics outside the phonon band is formed. When the phonon spectrum is inverted, the kink and antikink become mutually repulsive solitary waves with oscillatory tails, and their collision is possible only for velocities above a threshold value sufficient to overcome their repulsion.
\end{abstract}

\pacs{11.10.Lm, 11.27.+d, 05.45.Yv, 03.50.-z}
%11.10.Lm --- Nonlinear or nonlocal theories and models
%11.27.+d --- Extended classical solutions; cosmic strings, domain walls, texture,\\
%05.45.Yv --- Solitons,\\
%02.60.Cb --- Numerical simulation; solution of equations,\\
%02.30.Jr --- Partial differential equations,\\
%03.65.Pm --- Relativistic wave equations,\\
%03.50.-z --- Classical field theories,\\
%03.65.Ge --- Solutions of wave equations: bound states,\\
\maketitle

\section{Introduction}\label{sec:introduction}

Continuum and discrete Klein--Gordon type equations contribute to the understanding of many physical phenomena \cite{morris,BraunKivshar,Bookphi4,BookSGE}. In discrete models the Lorentz invariance is lost and several new important effects are observed, such as for example, the appearance of the Peierls--Nabarro potential, the associated reduction of soliton mobility, the radiation produced by moving solitary waves, etc.~\cite{Bookphi4,BookSGE,BraunKivshar,PNpot,Alfimov}. The case of strong discreteness is of particular interest and it is encountered in many applications, e.g., in the description of arrays of Josephson junctions~\cite{Zolotaryuk}, dissipative nonlinear discrete systems \cite{DissipativeKG}, dynamics of crowdions~\cite{Crowdions0,Crowdions1,Crowdions2,Crowdions3,Crowdions4} and  dislocations~\cite{Dislocations,Dislocations1,Dislocations2} in crystals, propagation of domain walls in magnetic materials~\cite{Magnetic}, motion of spring-mass chains \cite{SpringMass}, in the discussion of electric charge transport in molecular chains~\cite{ElectricCharge}. The consideration of the strongly discrete (anti-continuum) limit is a well-known approach aiming towards the analytical treatment of discrete breathers \cite{AC1,AC2}.

One particularly central problem in the dynamics of solitary waves is the analysis of their collision outcomes, especially beyond the merely-phase-shifting elastic wave interactions within integrable models~\cite{Campbell,CampbellPeyrard,anninos,Gani.PRE.1999,Collisions1,goodman,GH2007,Collisions2,Gani.PRD.2014,Gani.JHEP.2015,EGM2017,Moradi.CNSNS.2017,Moradi.JHEP.2017,Moradi.EPJB.2017,Bazeia.EPJC.2018,DSGE,DSGM,Belendryasova2019}. Continuum Klein--Gordon equations, apart from the famous integrable sine-Gordon example~\cite{BookSGE}, support exact solutions in the form of moving kinks which interact inelastically. Collisions between a kink and an antikink moving towards each other with initial velocities $\pm v_{\rm c}^{}$ have been extensively studied. It has been found that if $v_{\rm c}^{}>v^\ast$, where $v^\ast$ is a critical collision velocity, then the kink and antikink separate after the first collision~\cite{Campbell,CampbellPeyrard,anninos,Collisions1,goodman,GH2007}. On the other hand, collisions with $v_{\rm c}^{}<v^\ast$ produce a set of escape windows having fractal structure~\cite{anninos,Collisions1,goodman,GH2007}. When the collision velocity is within such a window,  the kink and antikink move away from each other after multiple collisions. The kinks' internal vibrational modes~\cite{Braun} have been extensively argued to be responsible for this effect: they store some energy of the kinks' translational motion which can be returned in the subsequent collisions, leading the kink and antikink to overcome their mutual attraction. If the kink and antikink do not split after a few collisions, they lose a substantial amount of energy to small-amplitude radiative wavepackets (emitted from the collision location) and, being unable to overcome the mutual attraction, create a bound oscillatory state called bion, whose main frequency lies below the phonon band but higher harmonics within the band. Resonating with the phonons, the bion constantly radiates energy and its amplitude gradually decreases.

The effect of weak discreteness on the solitary wave collisions was studied in Refs.~\cite{Collisions1,Collisions2} and radiationless energy exchange between colliding quasi-particles was described. On the other hand, the collision of solitary waves in strongly discrete Klein--Gordon systems has not been studied so far. The reason is the above mentioned immobility of kinks in the presence of Peierls--Nabarro potential induced by discreteness. However this difficulty can be overcome by considering exceptional discretizations \cite{BOP2005} of the Klein--Gordon equations where the static Peierls--Nabarro potential is precisely zero \cite{BOP2005,Speight1997,Speight1999,kevrekidis2003class,kevrekidis2005Asymptotic,dmitriev2006standard,dmitriev2005discrete,cooper2005exact,speight2006kinks,DKKS2007,KDS2009,BT1997,journal1}. In Ref.~\cite{roy2007comparative}, kink-antikink collisions have been analyzed in various models free of the static Peierls--Nabarro potential for the case of weak discreteness (lattice spacing $h\sim 0.1$). It has been found that collisions become more elastic with increasing $h$ because the critical collision velocity $v^*$, above which kinks separate after the first collision, reduces for larger $h$.

The absence of the static Peierls--Nabarro potential implies that the kink can move along the chain practically radiating no energy if its velocity is very small and its profile is not affected by the dynamical effects. Some discrete Klein--Gordon models support kinks or nanopterons moving with a permanent profile~\cite{SZE,SZ,OPB}, but such motion is observed only at so-called ``transparent points'', i.e., at isolated velocities at which kinks can propagate in a discrete system as traveling waves. In particular, in Ref.~\cite{OPB} three discrete $\phi^4$ equations free of the static Peierls--Nabarro potential were considered. The authors have shown that, in addition to the vanishing velocity, the discrete $\phi^4$ equation proposed by Speight and Ward~\cite{Speight1997,Speight1999} supports a single isolated velocity, the model of Kevrekidis~\cite{kevrekidis2003class} supports three velocities, and the model of Bender and Tovbis~\cite{BT1997} supports none such velocities.

In the present work, for the $\phi^4$ equation discretized according to the method proposed in Refs.~\cite{Speight1997,Speight1999}, we analyze kink-antikink collisions in the regime of strong discreteness ($h\sim 1$) in the absence of the static Peierls--Nabarro potential. We find that this setting is conducive to the emergence of multiple new features including the formation of persistent discrete breathers (rather than bions) in the strongly discrete regime. Another key finding is the potential  fundamental modification of the nature of the interaction from an attractive to a repulsive one upon the inversion of the phonon band for strong discreteness in such models.

Our presentation is structured as follows. In Section~\ref{sec:model} we present the (exceptional discretization) models and some of their principal static properties. Then in Section~\ref{sec:collisions} we examine the collisions of kinks and antikinks in these models. Finally, in Section~\ref{sec:conclusion} we summarize our findings and present our conclusions, as well as some challenges for future work.

\section{The $\phi^4$ field model and its exceptional discretization}\label{sec:model}

\subsection{Continuum $\phi^4$ equation and its conventional discretization}\label{sec:phi4model}

The Klein--Gordon field-theoretic model can be defined by the Hamiltonian
\begin{eqnarray} \label{eq:lagrangy}
H = \int_{-\infty}^{+\infty}\left[\frac{1}{2}\phi_t^2 + \frac{1}{2}\phi_x^2 + V(\phi)\right]dx,
\end{eqnarray}
where $\phi(x,t)$ is a real scalar function of spatial and temporal coordinates $x$ and $t$, respectively. The subscripts $x$ and $t$ denote differentiation with respect to the corresponding coordinate. The function $V(\phi)$ defines the on-site potential, which for the $\phi^4$ model reads:
\begin{eqnarray}
V(\phi) = \frac{1}{2}\left(1-\phi^2\right)^2.
\label{eq:Phi4Potential}
\end{eqnarray}
The Hamiltonian \eqref{eq:lagrangy} with the potential \eqref{eq:Phi4Potential} gives the following equation of motion:
\begin{eqnarray}
\phi_{tt}^{} - \phi_{xx}^{} - 2\phi\left(1-\phi^2\right) = 0,
\label{eq:Phi4EOM}
\end{eqnarray}
which can be transformed to the form of the $\phi^4$ model considered in Ref.~\cite{Speight1997} by rescaling the spatial and temporal coordinates by a factor of 1/2. This equation has an exact solution in the form of a moving kink (antikink),
\begin{eqnarray}
\phi=\pm\tanh\frac{x-x_0^{}-vt}{\sqrt{1-v^2}},
\label{eq:Phi4Solution}
\end{eqnarray}
for the upper (lower) sign, which propagates with the velocity $v$ starting at $t=0$ from the initial position $x=x_0^{}$. Substituting Eq.~\eqref{eq:Phi4Solution} into Eq.~\eqref{eq:lagrangy} one finds  for the total energy of the continuum kink
\begin{eqnarray}
E_{\rm K}^{\rm c} = \frac{4}{3\sqrt{1-v^2}}.
\label{eq:KinkEner}
\end{eqnarray}

The discrete $\phi^4$ equation is introduced on the lattice $x=nh$ with spacing $h>0$, where $n$ is integer. The conventional discretization of the $\phi^4$ equation \eqref{eq:Phi4EOM} reads \cite{Campbell}:
\begin{eqnarray}\label{eq:Classic}
\ddot{\phi}_n^{} = \frac{1}{h^2}\left(\phi_{n-1}^{} - 2\phi_n^{} + \phi_{n+1}^{}\right) + 2\phi_n^{}\left(1-\phi_n^2\right),
\end{eqnarray}
where $\phi_n^{}(t)=\phi(nh,t)$ and differentiation with respect to time is denoted by overdot. 
This discretization conserves the following Hamiltonian:
\begin{eqnarray} \label{H1}
   H = \frac{h}{2} \sum_n \left[ \dot{\phi}_n^2
   + \left(\frac{\phi_{n+1}^{}-\phi_n^{}}{h}\right)^2
   +\left(1 - \phi_n^2 \right)^2 \right].
\end{eqnarray}
Collision of highly discrete kinks cannot be studied within this model because the kinks are trapped by the Peierls--Nabarro potential and cannot propagate freely. 

\subsection{Exceptional discretization of the $\phi^4$ equation}\label{sec:discretephi4model}

Speight has derived the following discrete $\phi^4$ model \cite{Speight1997,Speight1999}:
\begin{eqnarray} \label{SpeightWardphi4}
\ddot{\phi}_n = \left(\frac{1}{h^2}+\frac{1}{3}\right)\left(\phi_{n-1}^{} - 2\phi_n^{} + \phi_{n+1}^{}\right)
\nonumber \\ + 2\phi_n^{} - \frac{1}{9}\left[2\phi_n^3 + \left(\phi_n^{}+\phi_{n-1}^{}\right)^3 + \left(\phi_n^{}+\phi_{n+1}^{}\right)^3\right] \equiv f_n^{},
\end{eqnarray}
which possesses the Hamiltonian
\begin{equation}
H = \frac{h}{2} \sum\limits_{n}\left(\dot{\phi}_n^2+u^2 \right),
\label{eq:H2}
\end{equation}
where
\begin{equation}
u\equiv \pm\frac{\phi_n^{}-\phi_{n-1}^{}}{h} - 1 + \frac{\phi_{n-1}^2+\phi_{n-1}^{}\phi_n^{}+\phi_n^2}{3}.
\label{eq:phi4twopoint}
\end{equation}
% u IS NOT 0 generally.  It is 0 for equilibrium states.
%
For equilibrium states, we have that $u=0$, which can be used for obtaining the kink profile (see below). Note that we have denoted the right-hand side of Eq.~\eqref{SpeightWardphi4} as $f_n^{}$, which is the force acting on $n$-th particle from the two neighboring particles and from the on-site potential.

It can be proved \cite{Speight1997,Speight1999} that static kinks of the model \eqref{SpeightWardphi4} can be derived iteratively from the two-point map by setting \eqref{eq:phi4twopoint} equal to $0$, which is a quadratic algebraic equation having the roots
\begin{eqnarray}
\phi_{n\pm1}^{} = -\frac{\phi_n^{}}{2} \mp \frac{3}{2h} \pm \frac{\sqrt{3}}{2}\sqrt{-\phi_n^2\pm\frac{6}{h}\phi_n^{}+\frac{3}{h^2}+4}.
\label{eq:Phi4kinksolution}
\end{eqnarray}
One can take in Eq.~\eqref{eq:Phi4kinksolution} either the upper or the lower signs. The iterations can be started from any initial value $|\phi_n^{}|<1$ to produce a static kink placed arbitrarily with respect to the lattice. The on-site kink is found for $\phi_n^{}=0$ and the inter-site kink for $\phi_n^{}=3/h-\sqrt{3+9/h^2}$. All such kinks have exactly the same potential energy and thus they do not experience a static Peierls--Nabarro potential. Kinks exist also in the particular case of $h=1$ and they can be found from the iterative formula \eqref{eq:Phi4kinksolution}.

\begin{figure}[t!]
\begin{center}
  \centering
  {\includegraphics[width=0.85
%  \textwidth]{fig1.eps}}
  \textwidth]{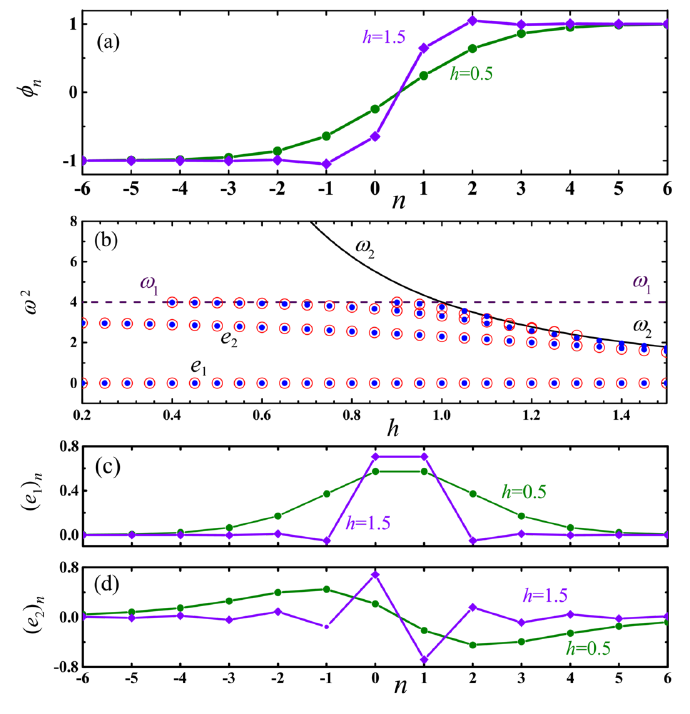}}
%  \hspace{1mm}
 \caption{(a) Inter-site static kink profiles for $h=0.5$ (dots) and $h=1.5$ (rhombuses). For $h<1$ the kink's tails are monotonic, while for $h>1$ they oscillate around the asymptotic states $\phi_n^{}=\pm 1$. (b) Borders of the phonon spectrum, $\omega_1^{}$ and $\omega_2^{}$, as functions of $h$, shown by the dashed and solid lines, respectively. The lines cross at $h=1$. The scattered data shows the frequencies of the small-amplitude vibrational modes localized on the on-site (dots) and inter-site (circles) static kinks. The zero-frequency mode $e_1^{}$ is the Goldstone translational mode, which is used for kink boosting. The second lowest frequency mode $e_2^{}$ is the kink's internal mode. (c), (d) Profiles of the Goldstone translational mode and kink's internal mode, respectively, for the kinks shown in (a).}
 \label{Fig:01}
\end{center}
\end{figure}

Examples of the inter-site static kink profiles, constructed by iterating Eq.~\eqref{eq:Phi4kinksolution}, are shown in Fig.~\ref{Fig:01}(a)
for $h=0.5$ by dots and for $h=1.5$ by rhombuses. Note that for $h<1$ the kink's tails are monotonic, while for $h>1$ they oscillate around the asymptotic states $\phi_n^{}=\pm 1$. The latter will play a key role in the modified interaction of the kink and antikink for $h>1$, as we will show below.

\begin{figure}[t!]
\begin{center}
  \centering
%  \subfigure[]{\includegraphics[width=0.45\textwidth]{fig2a.eps}}
  \subfigure[]{\includegraphics[width=0.45\textwidth]{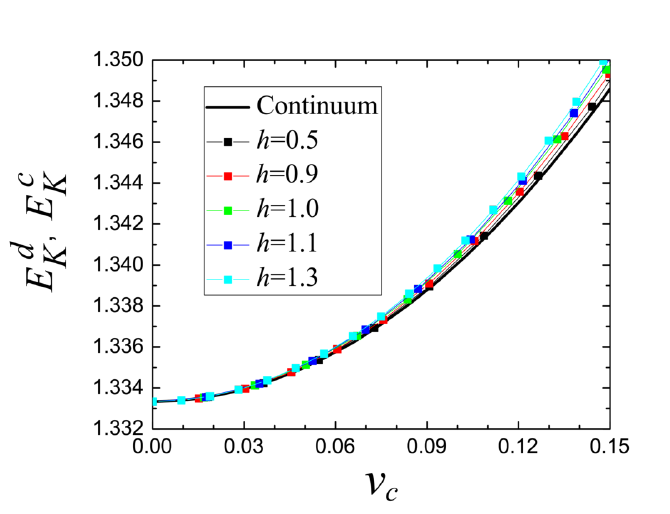}}
%  \hspace{1mm}
%  \subfigure[]{\includegraphics[width=0.45\textwidth]{fig2b.eps}}
  \subfigure[]{\includegraphics[width=0.45\textwidth]{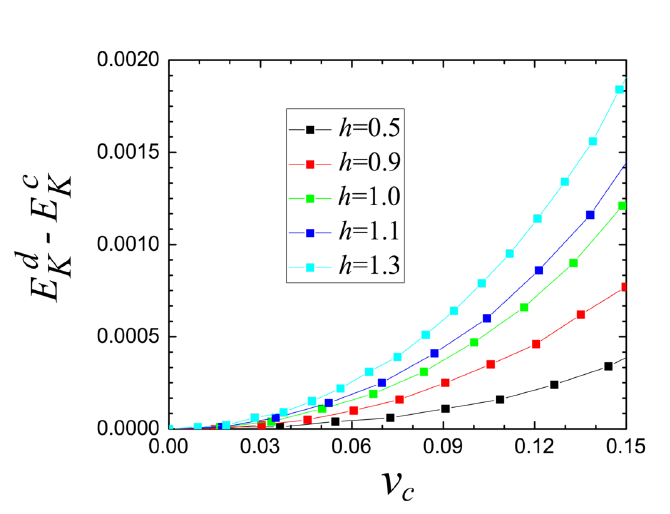}}
%  \hspace{1mm}
    \caption{(a) Total (potential plus kinetic) energy of the discrete kink, $E^{\rm d}_{\rm K}$, as a function of its velocity for different values of the discreteness parameter $h$. The solid line shows the prediction of the continuum theory, $E^{\rm c}_{\rm K}$, Eq.~\eqref{eq:KinkEner}. (b) The difference $E^{\rm d}_{\rm K}-E^{\rm c}_{\rm K}$ as a function of kink velocity for various values of the discreteness parameter $h$.}
  \label{Fig:02}
\end{center}
\end{figure}

Let us discuss the energy of the discrete kink. As it has been shown in the original paper by Speight \cite{Speight1997}, the energy of the static kink does not depend on the discreteness parameter $h$ and thus, it is equal to 4/3, as it follows from Eq.~\eqref{eq:KinkEner}, which is the result for the continuum kink ($h\rightarrow 0$). However, Speight’s theory says nothing about the kink dynamics and we have done numerical simulations to find the total (kinetic plus potential) energy of the discrete kink, $E^{\rm d}_{\rm K}$, for different values of $h$ and compared it to the prediction of the continuum theory, $E^{\rm c}_{\rm K}$, Eq.~\eqref{eq:KinkEner}. The result is presented in Fig.~\ref{Fig:02}(a),
%%%%%%%%%%%%%%%%%%%%%%%%%%  Fig. 2
where the solid line shows the prediction of the continuum theory, Eq.~\eqref{eq:KinkEner}, and the dots stand for various $h$ according to the legend. To better see the effect of $h$ on the energy of the moving discrete kink, in Fig.~\ref{Fig:02}(b) the difference $E^{\rm d}_{\rm K}-E^{\rm c}_{\rm K}$ is given as a function of $v_{\rm c}^{}$. It can be concluded that the energy of the moving discrete kink exceeds the energy predicted by the continuum theory and upon increasing $h$ the difference increases. The difference also increases with growing kink velocity $v_{\rm c}^{}$, but for vanishing kink velocity, kink energy is indeed equal to $4/3\approx 1.3333$ and it does not depend on $h$~\cite{Speight1997}. 

\subsection{Spectrum of vacuum and small-amplitude vibrations localized on the kink}\label{sec:Spectrum}

Let $\phi_n^0$ be an equilibrium static solution of Eq.~\eqref{SpeightWardphi4}. Small-amplitude oscillations around this solution can be studied by inserting $\phi_n^{}(t)=\phi_n^0+\varepsilon_n^{}(t)$ into the equation of motion \eqref{SpeightWardphi4}, where $\varepsilon_n^{} \ll 1$, and obtaining the linearized equation in the form 
\begin{eqnarray}
\ddot{\varepsilon}_n^{}&=&\frac{1}{h^2}\left(\varepsilon_{n-1}^{} -2\varepsilon_n^{} +\varepsilon_{n+1}^{}\right)+\frac{1}{3}\left[1-\left(\phi_n^0+\phi_{n-1}^0\right)^2\right]\varepsilon_{n-1}^{} \nonumber\\ &+&\frac{1}{3} \left[1-\left(\phi_n^0 +\phi_{n+1}^0\right)^2\right]\varepsilon_{n+1}^{}+\frac{1}{3}\left[4-2\left(\phi_n^0\right)^2-\left(\phi_n^0+\phi_{n-1}^0\right)^2-\left(\phi_n^0+\phi_{n+1}^0\right)^2\right]\varepsilon_n^{}.
\label{eq:Phi4Liniriazation}
\end{eqnarray}
Substituting into Eq.~\eqref{eq:Phi4Liniriazation} the ansatz $\varepsilon_n^{}=\exp(i q n -i \omega t)$, where $\omega$ is the frequency and $q$ is the wavenumber, one obtains the eigenvalue problem for finding the spectrum of small-amplitude vibrations around the static solution $\phi_n^0$.

In particular, the spectrum of the vacuum solution $\phi_n^0=\pm1$ is 
\begin{eqnarray}
\omega^2=4+4\frac{1-h^2}{h^2}\sin^2\left(\frac{q}{2}\right).
\label{eq:Phi4Spectrum}
\end{eqnarray}
This spectrum lies in between
\begin{eqnarray}
\omega_1^2 = 4 \quad {\rm and} \quad \omega_2^2 = 4 + 4\frac{1-h^2}{h^2}.
\label{eq:SpectrumBorders}
\end{eqnarray}
At $h=1$ one has $\omega_1^{}=\omega_2^{}$, i.e., the width of the spectrum vanishes and hence linear modes arise at a single frequency, namely $\omega=2$. This situation where the frequency is independent of wavenumber is referred to as a flat band and is of particular interest in recent studies~\cite{flachflat}. For $h<1$ the short waves ($|q|\approx \pi$) have frequencies higher than the long waves ($|q|\approx 0$). For $h>1$ the situation is reversed.

For the static kink solution $\phi_n^0$ the eigenvalue problem is solved numerically. For this, a kink is placed in the middle of the lattice of $N=200$ particles with boundary conditions $\phi_1^{}=-1$, $\phi_N^{}=1$. The solution of the eigenvalue problem gives $N-2$ eigenfrequencies $\omega_k$ and the same number of eigenvectors, $(e_k^{})_n^{}$. Most of the eigenfrequencies lie within the phonon spectrum of vacuum but a few of them are below the relevant band. This is demonstrated in Fig.~\ref{Fig:01}(b). The borders of the phonon spectrum as functions of $h$, Eq.~\eqref{eq:SpectrumBorders}, are shown by the dashed and solid lines. Dots and circles indicate frequencies of the modes localized on the on-site and inter-site kink, respectively. For any $h$ there exists the zero frequency mode $e_1^{}$, which is the translational (Goldstone) mode. The mode $e_2^{}$, which has the lowest non-zero frequency, is nothing but the kink's internal mode, which in the continuum limit ($h \rightarrow 0$) has frequency $\sqrt 3$. In Fig.~\ref{Fig:01}(c) we plot the Goldstone modes and in Fig.~\ref{Fig:01}(d) the kink's internal modes for the inter-site kinks shown in~(a), i.e., for the discreteness parameter $h=0.5$ (dots) and $h=1.5$ (rhombuses). 

\subsection{Interaction of well separated kink and mirror image antikink}\label{sec:ForceCalc}

\begin{figure}[t!]
\begin{center}
  \centering
  {\includegraphics[width=0.85
% \textwidth]{fig3.eps}}
 \textwidth]{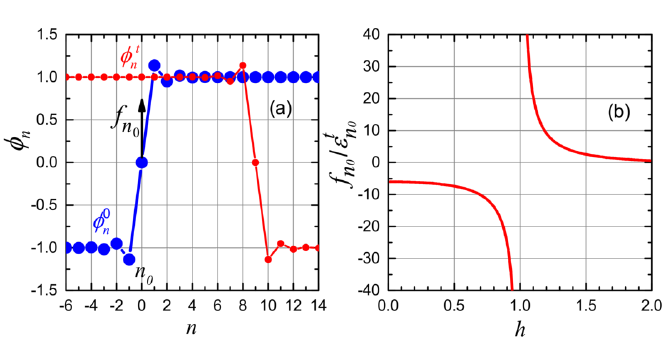}}
%  \hspace{1mm}
 \caption{(a) The setup of the calculation of the force acting on the central particle of the on-site kink from the antikink's tail. The exact on-site static kink solution $\phi_n^0$ (large circles) and the antikink (small circles) are shown. Here $h=2$ and hence the kinks have oscillatory tails. The kink is located at $n=n_0^{}=0$. (b) Normalized theoretically predicted force, $f_{n_0^{}}/\epsilon^t_{n_0^{}}$, acting on the central particle of the on-site kink from the antikink's tail as calculated from Eq.~\eqref{eq:fn0}.}
 \label{Fig:03}
\end{center}
\end{figure}

For further discussion it is instructive to understand how the kink and antikink interact with each other at large distances and how this interaction depends on the lattice spacing $h$. In the continuum models, the force acting between the kinks is usually calculated as minus gradient of the potential energy of their interaction. In the regime of strong discreteness ($h\sim 1$), the kink is localized on a few particles. Using this fact, we just calculate the force acting on the central particle of the on-site kink from the tail of an on-site antikink, which is located at some distance to the right of the kink, see Fig.~\ref{Fig:03}(a).
%%%%%%%%%%%%%%%%%%%%%%%%%%  Fig. 3
The high symmetry of the considered structure allows to solve the problem. The calculation will be done in the adiabatic approximation assuming that the kink and the antikink are at rest. Let $\phi_n^0$ be the static kink solution and $\phi_n^t$ be the antikink tail. We assume the tail solution to be of the form (as will be justified later)
\begin{equation}
   \phi_n^t=1+\epsilon^t_n,
   \label{eq:ans}
\end{equation}
where $|\epsilon_n^{t}| \ll 1$. The force acting on the $n$-th site, which is within the kink, can be calculated by substituting the linear superposition $\phi_n=\phi_n^0+\epsilon^t_n$ into Eq.~\eqref{SpeightWardphi4} and linearizing with respect to $\epsilon^t_n$. The result is
\begin{eqnarray}
f_n^{}&=&\frac{1}{h^2}\left(\epsilon^t_{n-1} - 2\epsilon^t_n + \epsilon^t_{n+1}\right) + \frac{1}{3} \left[1 - \left(\phi_n^0+\phi_{n-1}^0\right)^2 \right] \epsilon^t_{n-1} + \frac{1}{3} \left[1 - \left(\phi_n^0 +\phi_{n+1}^0\right)^2 \right] \epsilon^t_{n+1}
\nonumber\\ &+& \frac{1}{3} \left[4 - 2\left(\phi_n^0\right)^2 - \left(\phi_n^0+\phi_{n-1}^0\right)^2 - \left(\phi_n^0+\phi_{n+1}^0\right)^2 \right] \epsilon_n^t.
\label{eq:Phi4LiniriazationTail}
\end{eqnarray}
Notice that linear combinations of the kink solution and the antikink tail solution considered here can be used for kinks with short-range tails, as in our case, but this approximation may not work for kinks with long-range tails, see, e.g., Refs.~\cite{Belendryasova2019,Christov.PRD.2019,Christov.PRL.2019,Manton2019,Khare.JPA.2019}.

An approximate kink tail solution can be derived by substituting Eq.~\eqref{eq:ans} into Eq.~\eqref{eq:Phi4kinksolution} and expanding with respect to $\epsilon^t_n$ up to the second power. The resulting iterative formula reads 
\begin{equation}   
   \epsilon^t_{n+1}=\frac{1-h}{1+h}\epsilon^t_n-\frac{h(3+h^2)}{3(1+h)^3}(\epsilon^t_n)^2.
   \label{eq:tail}
\end{equation} 
It can be seen from the linear term of Eq.~\eqref{eq:tail} that for $h<1$ all $\epsilon^t_n$ have the same sign, so that the kink's tail monotonously approaches the value $\phi_n^{}=1$. For $h>1$ the tail oscillates near $\phi_n^{}=1$ since $\epsilon^t_n$ and $\epsilon^t_{n+1}$ have opposite signs. If $h=1$, the linear term in the expansion \eqref{eq:tail} vanishes and we have a purely anharmonic chain with a corresponding short-range kink tail. Note that the antikink tail is a mirror image of the kink tail \eqref{eq:tail}, that is why the antikink tail can be described using
\begin{equation}
   \epsilon^t_{n-1}=\frac{1-h}{1+h}\epsilon^t_n-\frac{h(3+h^2)}{3(1+h)^3}(\epsilon^t_n)^2.
   \label{eq:AKtail}
\end{equation} 

The ansatz \eqref{eq:ans} can be justified by considering the vacuum solution $\phi_n^0=1$ for all $n$ and getting from Eq.~\eqref{eq:Phi4Liniriazation} the following static equation for the small deviation from the vacuum: $(1/h^2-1)(\varepsilon_{n-1}^{}-2\varepsilon_n^{}+\varepsilon_{n+1}^{})-4\varepsilon_n^{}=0$. Looking for the solution to this equation in the form $\varepsilon_n^{}=q^n$ one obtains the quadratic characteristic equation having the roots $q_{1,2}^{}=(1 \mp h)/(1\pm h)$, which is equivalent to the linear part of kink tail solution \eqref{eq:tail}.

The force $f_n^{}$ acting on any site within the kink can be now calculated for any given value of $\epsilon^t_n \ll 1$ after finding $\epsilon^t_{n\pm 1}$ from Eq.~\eqref{eq:AKtail} and substituting into Eq.~\eqref{eq:Phi4LiniriazationTail}. Recall that the static kink solution can be found iteratively for any given $-1<\phi_n^0<1$ from Eq.~\eqref{eq:Phi4kinksolution}.

The analytical expression for the force $f_n^{}$ can be simplified for the case of the highly symmetric on-site kink. Let the on-site kink be located at $n=n_0^{}$, i.e., $\phi_{n_0^{}}^0=0$. Then from Eq.~\eqref{eq:Phi4kinksolution} one finds the displacements for the neighboring sites:
\begin{equation}
   \phi_{n_0^{}\pm 1}^0=\mp\frac{3}{2h} \pm \frac{\sqrt{3}}{2}\sqrt{\frac{3}{h^2}+4}.
   \label{eq:phi0n0}
\end{equation}
For a given small value $\epsilon^t_{n_0^{}}$ we find $\epsilon^t_{n_0^{}\pm 1}$ from the linear part of Eq.~\eqref{eq:AKtail} and substituting these values together with $\phi_{n_0^{}}=0$ and $\phi_{n_0^{}+1}=-\phi_{n_0^{}-1}$ into Eq.~\eqref{eq:Phi4LiniriazationTail}, we obtain the force acting from the antikink's tail on the central particle of the on-site kink:
\begin{eqnarray}
   f_{n_0^{}}&=&-\frac{4\epsilon^t_{n_0^{}}}{(1-h)(1+h)} \nonumber \\
   &-&\frac{\epsilon^t_{n_0^{}}}{3}\left\{\frac{2\left(1+h^2\right)}{(1-h)(1+h)}\left[1-\left(\phi_{n_0^{}+1}^0\right)^2\right]-2\left(\phi_{n_0^{}+1}\right)^2+4\right\},
   \label{eq:fn0}
\end{eqnarray}
where $\phi_{n_0^{}+1}$ is given by Eq.~\eqref{eq:phi0n0}. Note that for the considered here kink and mirror image antikink the sign of $\epsilon^t_{n_0^{}}$ in Eq.~\eqref{eq:fn0} does not change. There are two possible interpretations of the displacements of particles $\phi_n^{}$, i.e., one can think of either longitudinal or transverse particle motion. In Fig.~\ref{Fig:03}(a) the force acting on the $n_0^{}$-th particle is shown in vertical direction because the particle displacements $\phi_n^{}$ are also shown in transverse direction. For positive such forces, the particle is moving upwards which means that a preceding particle is moving upward as well, leading the coherent structure to move to the left. By a symmetric argument, when the relevant force is negative, the waves will move toward each other. In this way, one can connect the force on individual particles to the force resulting on the nonlinear wave.

From Eq.~\eqref{eq:fn0} it can be seen that $ f_{n_0^{}}$ is proportional to $\epsilon^t_{n_0^{}}$. The dependence of $f_{n_0^{}}/\epsilon^t_{n_0^{}}$ on $h$ is shown in Fig.~\ref{Fig:03}(b). It is interesting to note that the sign of the force changes at $h=1$ so that the kink and antikink attract each other for $h<1$ and repel each other for $h>1$. This fact will be confirmed numerically in Sec.~\ref{sec:collisions}.

\begin{figure}[t!]
\begin{center}
  \centering
  {\includegraphics[width=0.6
% \textwidth]{fig4.eps}}
 \textwidth]{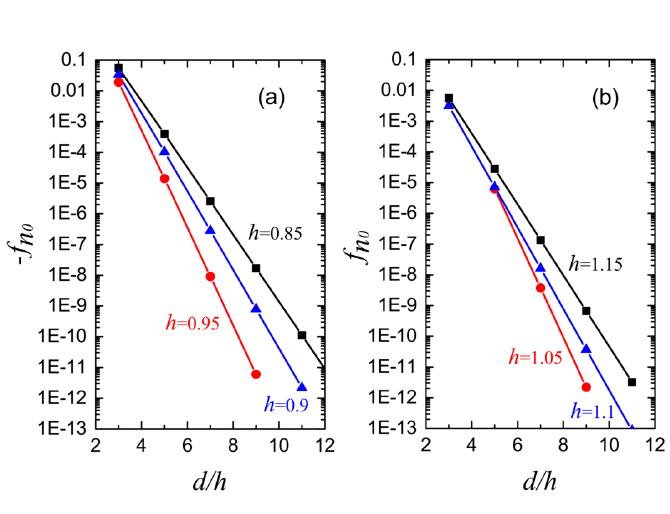}}
%  \hspace{1mm}
\caption{Force acting on the central particle of the on-site kink from the tail of the on-site antikink at the distance $d=nh$ from the kink: (a) the case of attractive interaction when $h<1$ and (b) the case of repulsive interaction $h>1$. The values of the lattice spacing $h$ are given for each curve.}
 \label{Fig:04}
\end{center}
\end{figure}

Note that the ratio $f_{n_0^{}}/\epsilon^t_{n_0^{}}$ diverges at $h=1$, see Fig.~\ref{Fig:03}(b). The values of the force $f_{n_0^{}}$ remain finite even for $h=1$. In Fig.~\ref{Fig:04}
%%%%%%%%%%%%%%%%%%%%%%%%%%%% Fig. 4
the values of the force $f_{n_0^{}}$ acting on the central particle of the on-site kink from the tail of the mirror image antikink are presented for the lattice spacing close to 1 as a function of the distance between the kink and antikink $d$ normalized by the lattice spacing $h$. In (a) $h<1$ and the force is negative since the kink and antikink attract each other. In (b) $h>1$ and the interaction between the kink and antikink is repulsive. The use of the logarithmic scale for the ordinate reveals exponential decay of the force with the distance between kink and antikink and this decay is faster for $h$ closer to 1. 

\section{Kink-antikink collisions}\label{sec:collisions}

\subsection{Simulation setup}\label{sec:SetupBoosting}

The set of equations of motion \eqref{SpeightWardphi4} was integrated numerically using the St\"ormer method of the sixth order~\cite{Bakhvalov} with the time step $\tau=0.005$. This method is efficient in finding a solution to the Cauchy problem for a system of second-order ordinary differential equations not containing the first derivative of the unknown function. The kink-antikink collisions were investigated in the chain having sufficiently large number of particles $N$, so that the radiation emitted by the colliding kinks does not reach the ends of the chain by the end of the simulation run. This way the effect of the radiation on the kink dynamics is avoided. Fixed boundary conditions were employed, $\phi_{\rm L}^{}=\phi_{\rm R}^{}=-1$, where $\phi_{\rm L}^{}$ and $\phi_{\rm R}^{}$ are the coordinates of the particles at the left and right ends of the chain, respectively (though the type of boundary conditions is not important for such sufficiently long chains). In typical runs we took the number of lattice points $N=2000$, which was sufficient to avoid the effect of the radiation reflected from the fixed ends of the chain on the kink-antikink collisions.

The moving kink can be obtained by using the Goldstone translational mode $e_1^{}$, see Fig.~\ref{Fig:01}(c). The initial conditions were formulated as follows. At $t=0$ we set $\phi_n^{}=\phi_n^0$, where $\phi_n^0$ is the static kink solution, and at $t=\tau$ the Goldstone mode is added, $\phi_n^{}=\phi_n^0+\delta(e_1^{})_n^{}$, with a small coefficient $\delta$ which defines the speed of the boosted kink. The eigenvector $e_1^{}$ is assumed to be normalized, $||e_1^{}||=1$. The resulting kink velocity is measured numerically and it is called collision velocity $v_{\rm c}^{}$.

The obtained kink moving with the velocity $v_{\rm c}^{}$ collides with its mirror image antikink having velocity $-v_{\rm c}^{}$ in the middle of the chain. The initial distance between the kink and antikink is taken sufficiently large for their exponential tails to not overlap.

\subsection{Numerical results}

\begin{figure}[t!]
\begin{center}
  \centering
  {\includegraphics[width=1.0
% \textwidth]{fig5.eps}
 \textwidth]{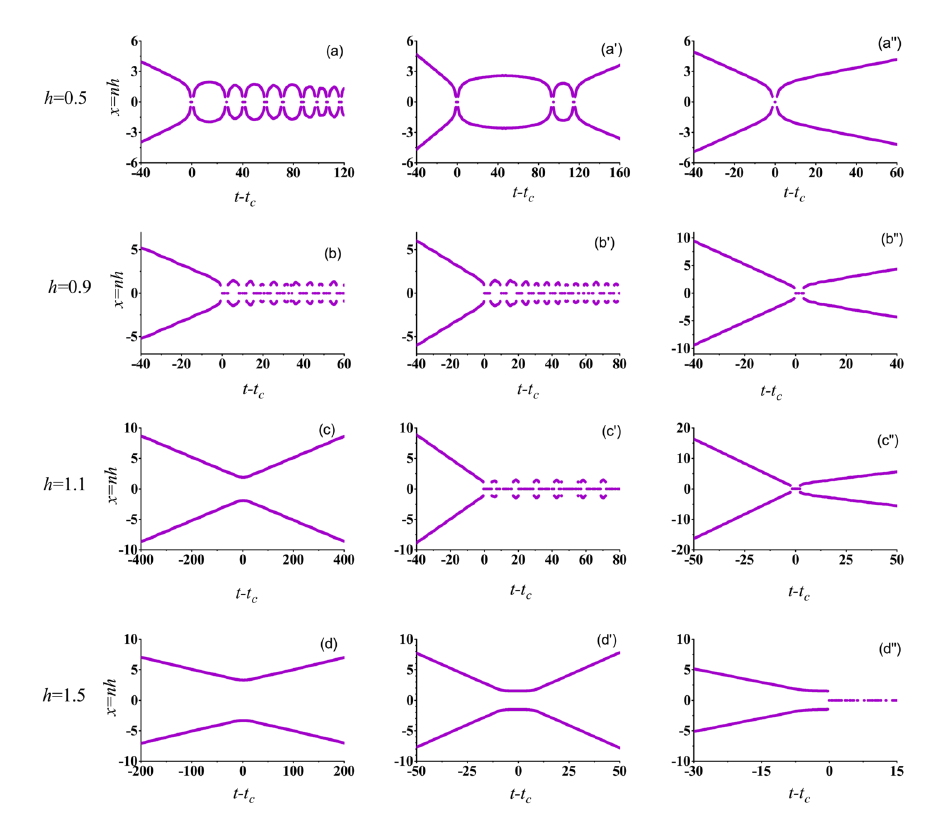}
 \label{fig:All9Fig}}
%  \hspace{1mm}
 \caption{Trajectories of colliding kinks and antikinks in the time-space plane for $h=0.5$ in (a)-(a$^{\prime\prime}$), for $h=0.9$ in (b)-(b$^{\prime\prime}$), for $h=1.1$ in (c)-(c$^{\prime\prime}$), and for $h=1.5$ in (d)-(d$^{\prime\prime}$),  shown by plotting the particles with the maximal energy. The origin of the temporal coordinate is chosen at the collision time moment $t_{\rm c}^{}$. The collision velocity $v_{\rm c}^{}$ increases in each row from the left to the right having the values: (a)~0.06062, (a$^{\prime}$)~0.08237, (a$^{\prime\prime}$)~0.09069; (b)~0.1, (b$^{\prime}$)~0.15, (b$^{\prime\prime}$)~0.2; (c)~0.01751, (c$^{\prime}$)~0.1958, (c$^{\prime\prime}$)~0.2501; (d)~0.01994, (d$^{\prime}$)~0.1481, (d$^{\prime\prime}$)~0.1486.}
 \label{Fig:05}
\end{center}
\end{figure}

\begin{figure}[t!]
\begin{center}
  \centering
  \subfigure[$\,h=0.5$, $v_{\rm c}^{}=0.06062$, cf.\ Fig.~\ref{Fig:05}(a). ]{\includegraphics[width=0.45\textwidth]{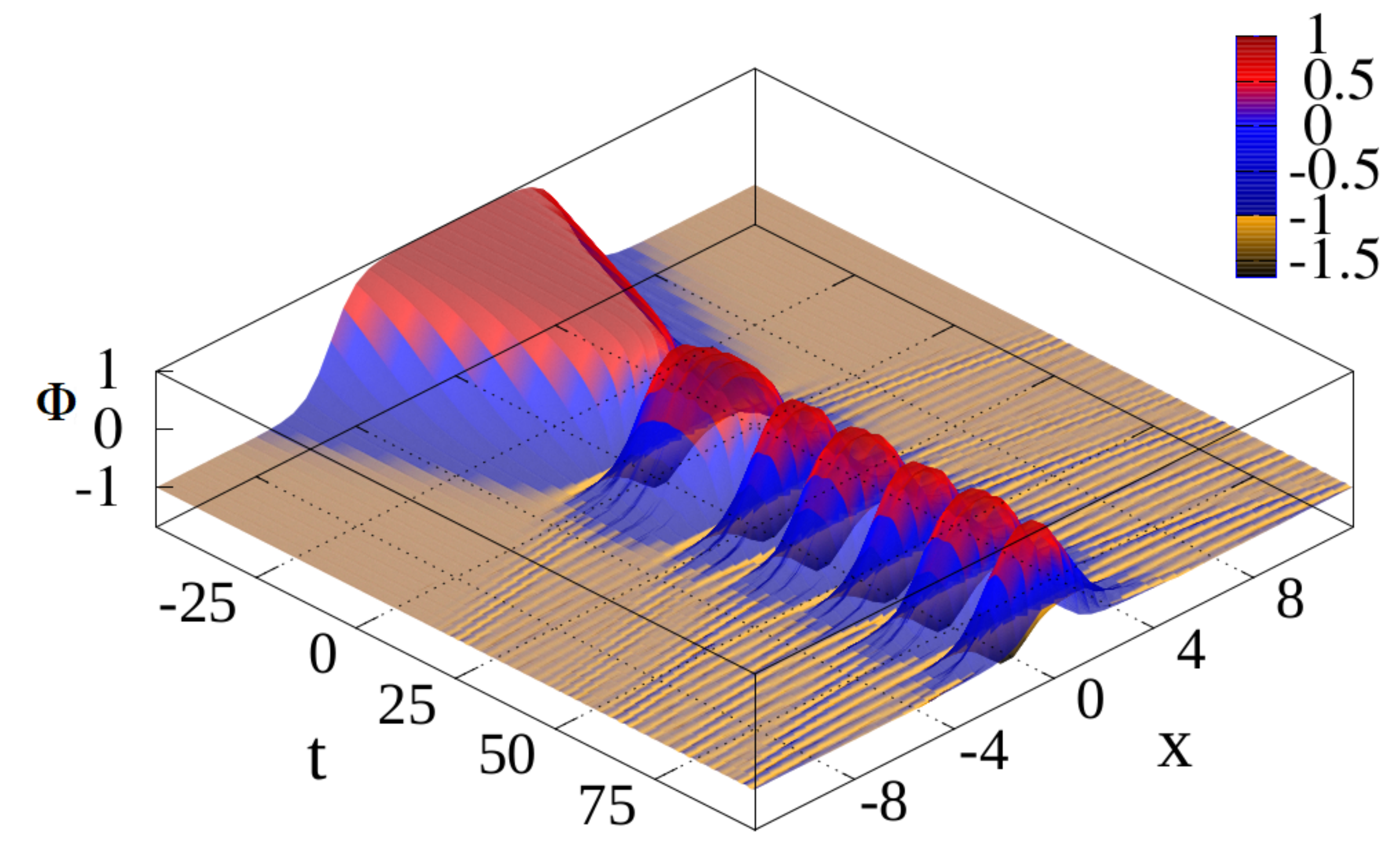}\label{fig:fig4a}}
%  \hspace{1mm}
  \subfigure[$\,h=0.5$, $v_{\rm c}^{}=0.08237$, cf.\ Fig.~\ref{Fig:05}(a').]{\includegraphics[width=0.45\textwidth]{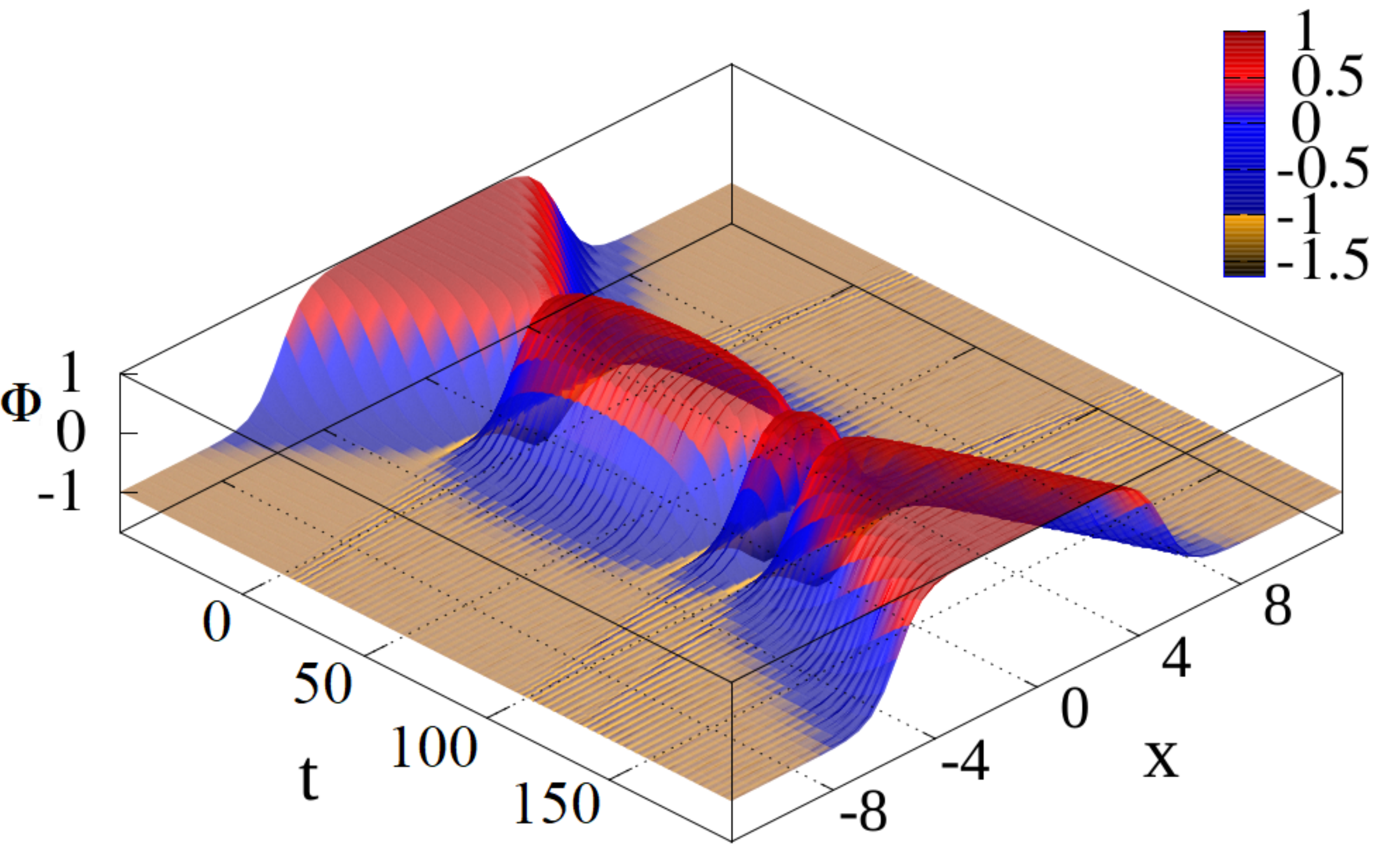}\label{fig:fig4a'}}
%  \hspace{1mm}
  \\
  \subfigure[$\,h=0.9$, $v_{\rm c}^{}=0.15$, cf.\ Fig.~\ref{Fig:05}(b').]{\includegraphics[width=0.45\textwidth]{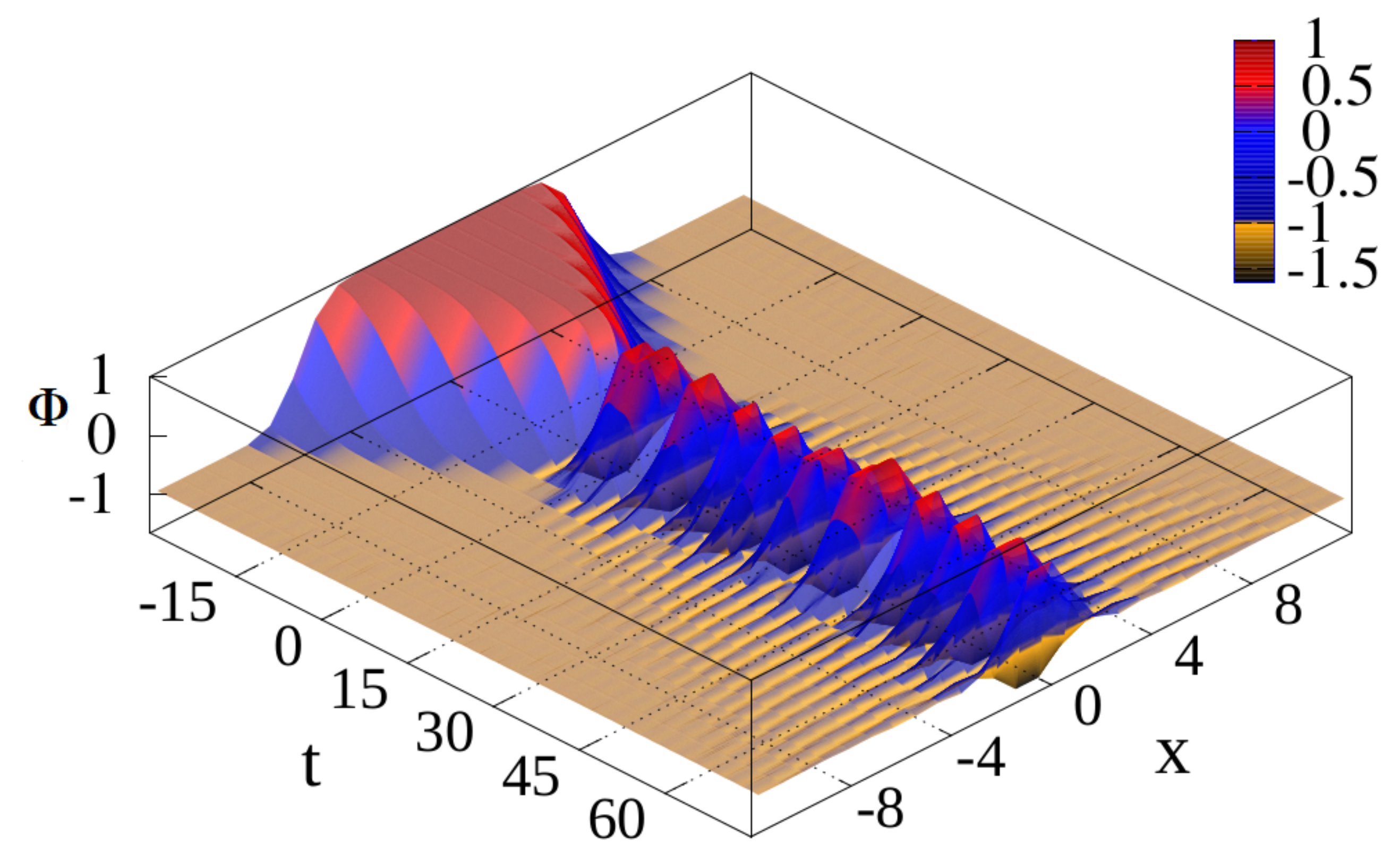}\label{fig:fig4b'}}
%  \hspace{1mm}
  \subfigure[$\,h=1.1$, $v_{\rm c}^{}=0.1958$, cf.\ Fig.~\ref{Fig:05}(c').]{\includegraphics[width=0.45\textwidth]{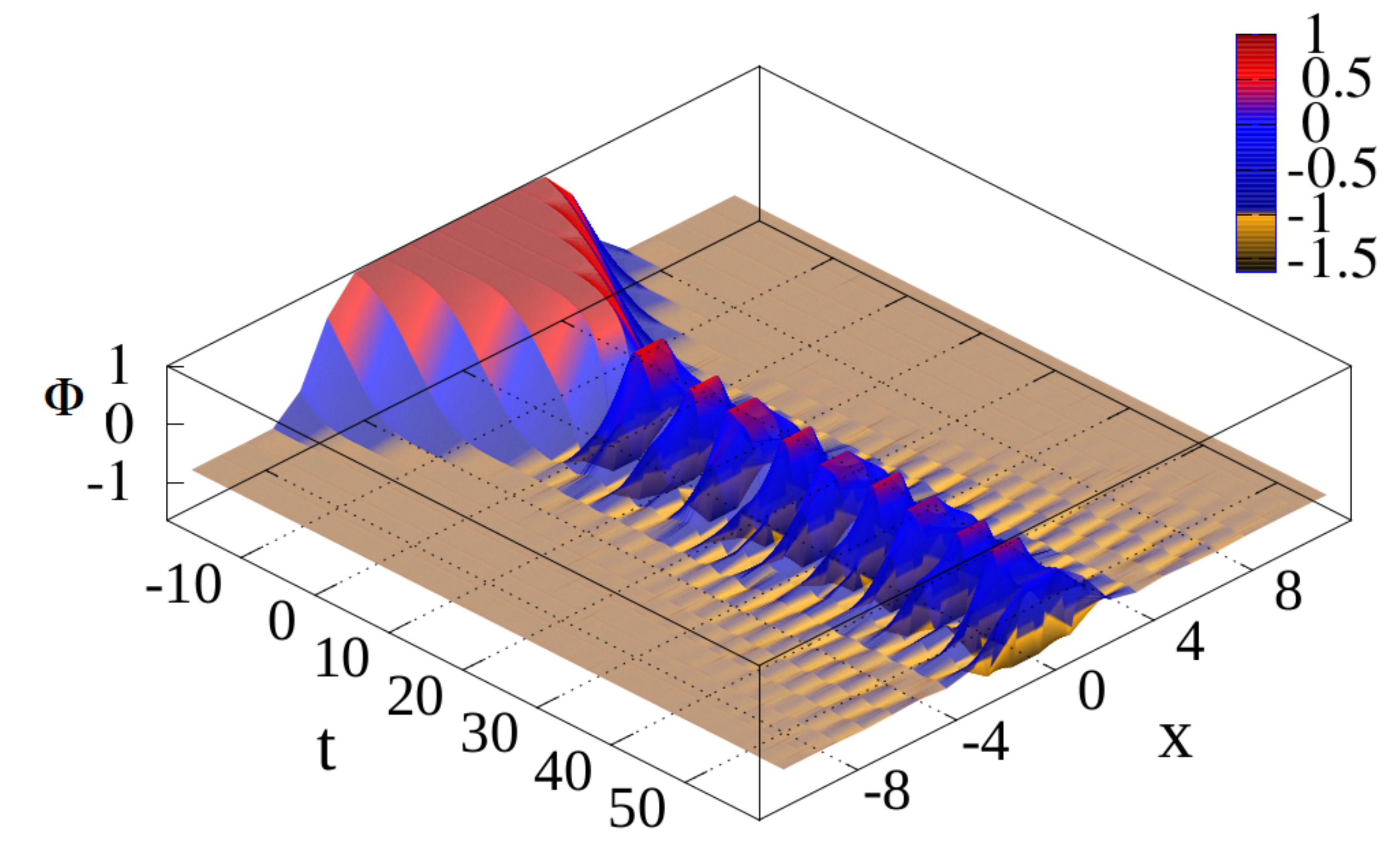}\label{fig:fig4c'}}
%  \hspace{1mm}
  \\
 \subfigure[$\,h=1.5$, $v_{\rm c}^{}=0.01994$, cf.\ Fig.~\ref{Fig:05}(d).]{\includegraphics[width=0.45\textwidth]{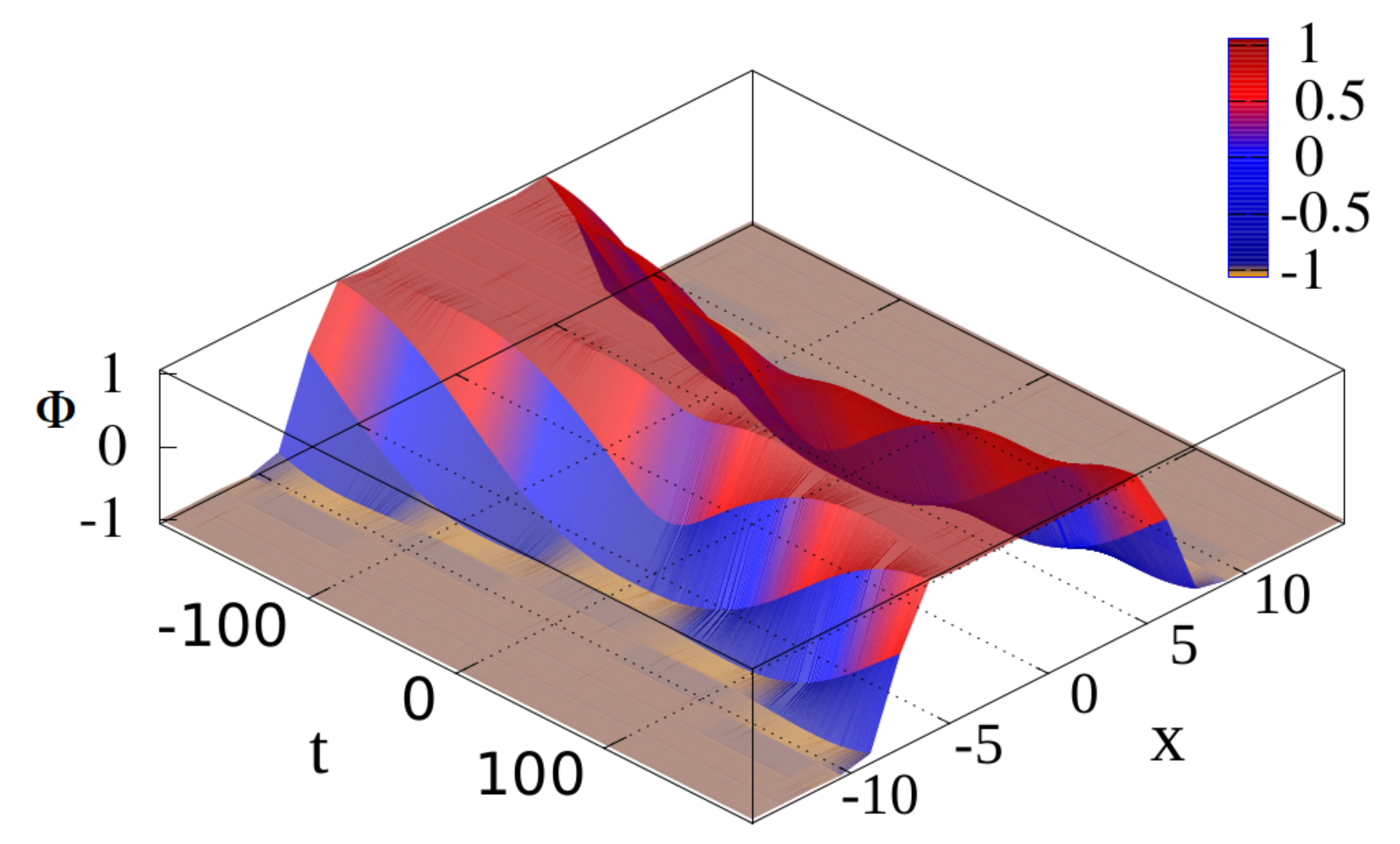}\label{fig:fig4d}}
%  \hspace{1mm}
 \subfigure[$\,h=1.5$, $v_{\rm c}^{}=0.1486$, cf.\ Fig.~\ref{Fig:05}(d").]{\includegraphics[width=0.45\textwidth]{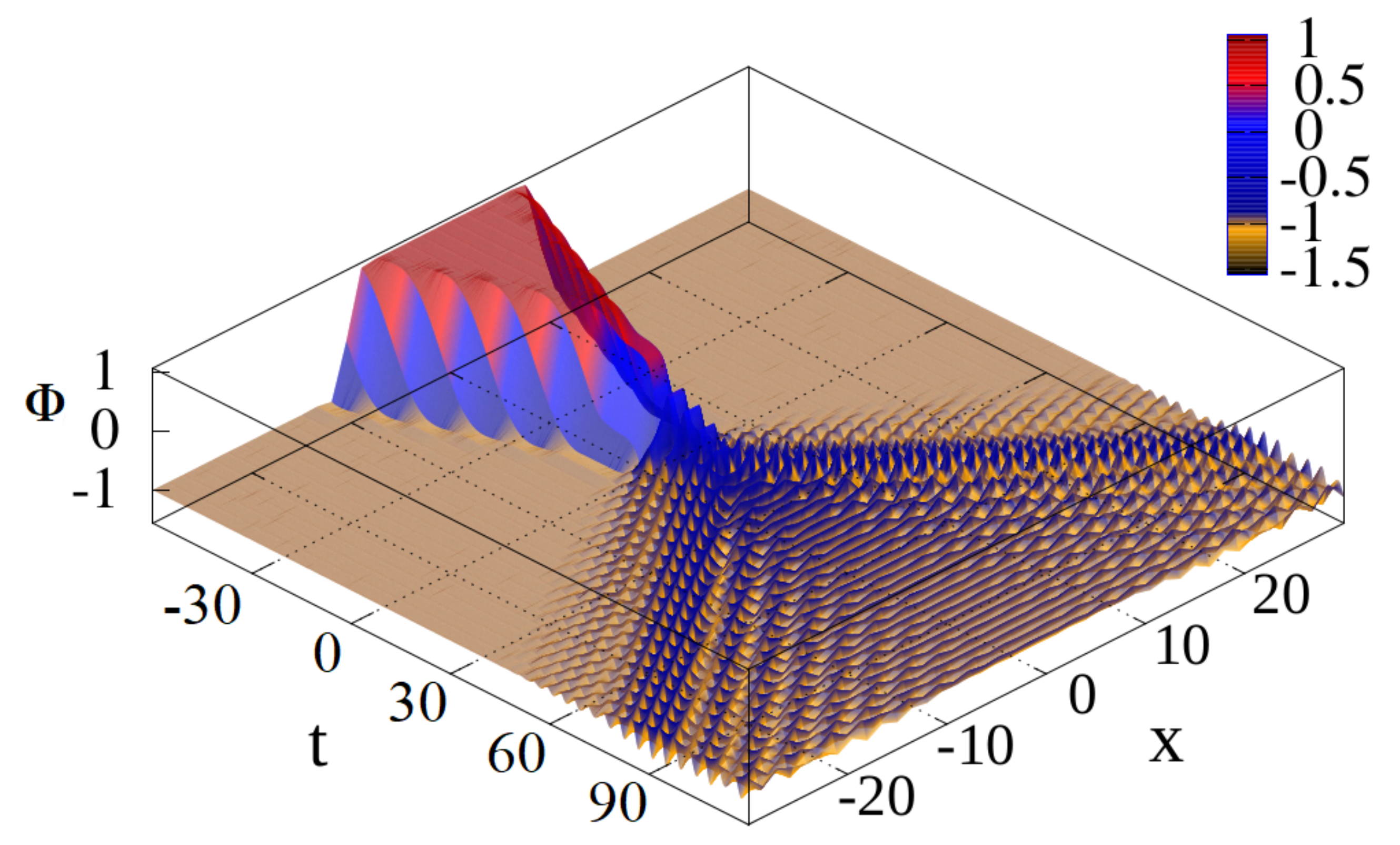}\label{fig:fig4d"}}
%  \hspace{1mm}
  \caption{3D plots showing kink-antikink collisions. For each panel the values of the discreteness parameter $h$ and collision velocity $v_{\rm c}^{}$, as well as the link to corresponding panel of Fig.~\ref{Fig:05} are provided.}
  \label{Fig:06}
\end{center}
\end{figure}

Examples of the kink-antikink collisions are presented in Fig.~\ref{Fig:05}: in (a)--(a$^{\prime\prime}$) for $h=0.5$, in (b)--(b$^{\prime\prime}$) for $h=0.9$, in (c)--(c$^{\prime\prime}$) for $h=1.1$, and in (d)--(d$^{\prime\prime}$) for $h=1.5$, by plotting the particles with the maximal energy.
%%%%%%%%%%%%%%%%%%%%%%%%%%%%% Fig. 5
The collision velocity in each row increases from the left to the right. Corresponding 3D plots are given in Fig.~\ref{Fig:06}.
%%%%%%%%%%%%%%%%%%%%%%%%% Fig. 6

It can be clearly inferred that the collisions are qualitatively different for $h<1$ and $h>1$, since the kink profiles are different in these two cases, see Eq.~\eqref{eq:tail} and Fig.~\ref{Fig:01}(a) for the kink tails, and the sign of the force acting between the kink and antikink changes at $h=1$, as it was demonstrated in Sec.~\ref{sec:ForceCalc}. Indeed, in line with the suggestive theoretical analysis of the previous section for $h<1$ ($h>1$) the kink and antikink attract (repel) each other.

At $h<1$ (two upper rows of Fig.~\ref{Fig:05}) the kink and antikink attract each other as in the continuum case. In the first row the phonon band width is relatively large and it is small in the second row. The collisions are not elastic and a part of the energy of the kink and antikink related to their translational motion is converted into  long-lived vibrational modes, while a smaller portion of energy is radiated in the form of small-amplitude waves. If the collision velocity is above a threshold value $v^\ast$, the coherent structures pass through each other and continue their motion with a reduced velocity $v<v_{\rm c}^{}$, as exemplified in (a$^{\prime\prime}$) and (b$^{\prime\prime}$). Plots (a) and (a$^{\prime}$) show collisions with velocity $v_{\rm c}^{}<v^\ast$. Here the waves after the collision cannot overcome their mutual attraction and collide again. In Fig.~\ref{Fig:05}(a) a bion is formed [see also 3D plot in Fig.~\ref{Fig:06}(a)], which is a kink-antikink bound state. The bion's frequency is below the phonon band, $\omega_{\rm B}^{}<\omega_1^{}=2$. From Eq.~\eqref{eq:SpectrumBorders}, for $h=0.5$ the upper edge of the phonon band is $\omega_2^{}=4$. This means that bion's higher harmonics are always within the phonon band, and thus the bion  radiates its energy due to the relevant resonance mechanism. In~Fig.~\ref{Fig:05}(a$^{\prime}$), we have a three-bounce collision after which the kinks separate and continue their motion with a velocity $v<v_{\rm c}^{}$ [also shown in Fig.~\ref{Fig:06}(b)]. Such multi-bounce collisions with $v_{\rm c}^{}<v^\ast$ are possible because the energy stored by the kinks' internal modes can be transformed back into energy of the kinks' translational motion. This suggests that here we are still in a regime proximal to the continuum limit where such phenomenology is well-known \cite{anninos,goodman,GH2007}. For a discrete breather (DB) to exist on top of the nonzero background, it is necessary that its frequency and all higher harmonics are outside of the phonon spectrum~\cite{AC1}. In Fig.~\ref{Fig:05}(b) and (b$^{\prime}$) the collision velocity is also below $v^*$. However in this case the width of the phonon band is small and a DB is formed with its frequency and all higher harmonics outside the phonon band. Formation of DBs can also be seen in the 3D plots, Fig.~\ref{Fig:06}(c,d). Thus, there is no mechanism for the breather to radiate its energy and it can be expected to persist over a long-time evolution.

For the analysis of the kink-antikink collisions in the case of $h>1$ one should take into account the kinetic energy of the interacting kinks which can help to overcome their mutual repulsion. This repulsion is clearly seen in Fig.~\ref{Fig:05}(c), (d) and (d$^{\prime}$), where the collision velocity is below a threshold value $v^{\ast\ast}$ sufficient to overcome their repulsion. In Fig.~\ref{Fig:05}(c$^{\prime}$), (c$^{\prime\prime}$) and (d$^{\prime\prime}$) the collision velocity is above $v^{\ast\ast}$ and the kinks indeed collide. A 3D picture of kink-antikink repulsion is presented in Fig.~\ref{Fig:06}(e). In Fig.~\ref{Fig:05}(c$^{\prime\prime}$) the kink and antikink emerge after the collision with a velocity smaller than $v_{\rm c}^{}$. In Fig.~\ref{Fig:05}(d$^{\prime\prime}$) [also in Fig.~\ref{Fig:06}(f)], as a result of the collision, a bion is produced with the main frequency within the relatively wide phonon band. The bion disappears after a few oscillations producing a burst of radiation. On the other hand, in Fig.~\ref{Fig:05}(c$^{\prime}$) [also in Fig.~\ref{Fig:06}(d)] the phonon band is narrow and a DB is formed with frequency $\omega_{\rm DB}^{}<\omega_2^{}$, i.e., below the phonon band with all higher harmonics above the phonon band, hence the relevant waveform is expected to persist, as is indeed
observed in the corresponding evolution dynamics.

\begin{figure}[t!]
\begin{center}
  \centering
  {\includegraphics[width=0.65
% \textwidth]{fig7.eps}}
 \textwidth]{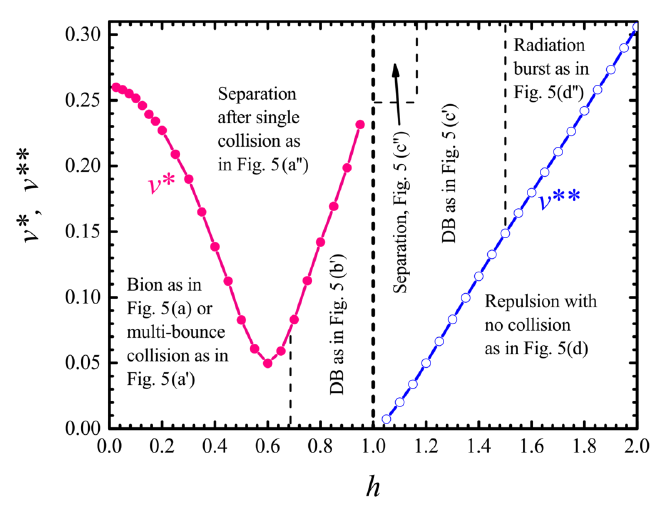}}
%  \hspace{1mm}
 \caption{Critical collision velocities $v^*$ and $v^{**}$ as functions of $h$, for $h<1$ and $h>1$, respectively. For $v_{\rm c}^{}>v^*$ kink and antikink separate after the first collision, as shown in Fig.~\ref{Fig:05}(a$^{\prime\prime}$). For $v_{\rm c}^{}<v^*$ and $h<0.7$ they either create a bion, Fig.~\ref{Fig:05}(a), or separate after a multi-bounce collision, Fig.~\ref{Fig:05}(a$^{\prime}$). For $v_{\rm c}^{}<v^*$ and $0.7<h<1$ a DB is formed, Fig.~\ref{Fig:05}(b) or (b$^{\prime}$). When $v_{\rm c}^{}>v^{**}$ the kink and antikink collide forming a DB ($1<h<1.5$, Fig.~\ref{Fig:05}(c$^{\prime}$)) or a burst of radiation ($h>1.5$, Fig.~\ref{Fig:05}(d$^{\prime\prime}$)). For sufficiently large $v_{\rm c}^{}$ and $h$ close to 1, the kink and antikink separate after the first collision, Fig.~\ref{Fig:05}(c$^{\prime\prime}$). For $v_{\rm c}^{}<v^{**}$ they repel each other, Fig.~\ref{Fig:05}(c) or (d), or (d$^{\prime}$).}
 \label{Fig:07}
\end{center}
\end{figure}

Two critical velocities were defined above, $v^*$ for $h<1$ and $v^{**}$ for $h>1$. In Fig.~\ref{Fig:07} the critical velocities are plotted as  functions of $h$.
%%%%%%%%%%%%%%%%%%%%%%%%%%%  Fig. 7
The critical velocity $v^*$ has a minimum at $h=0.6$; here, collisions are most elastic. $v^*$ denotes the threshold above which separation of the kink-antikink pair occurs after a single collision. The critical velocity $v^{**}$ increases linearly with $h$. $v_{\rm c}^{}<v^{**}$ denotes the scenario of repulsion without collision in the case of $h>1$. Depending on $h$ and the collision velocity, Fig.~\ref{Fig:07} is divided into parts where different collision scenarios are observed, as linked to the panels of Fig.~\ref{Fig:05}.

\begin{figure}[t!]
\begin{center}
  \centering
  {\includegraphics[width=0.65
% \textwidth]{fig8.eps}}
 \textwidth]{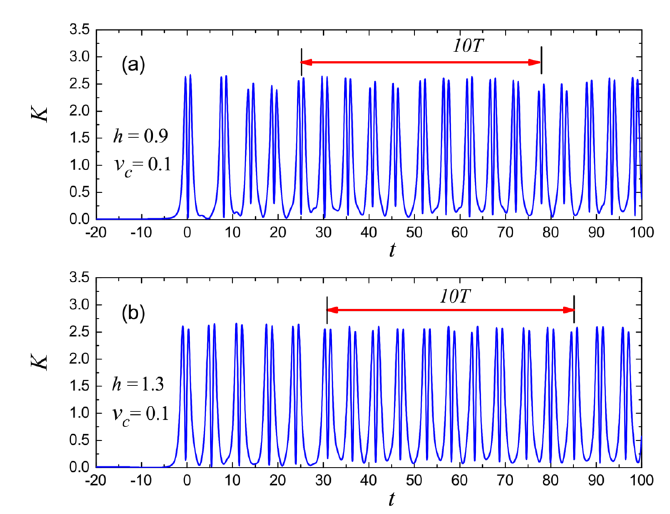}}
%  \hspace{1mm}
 \caption{Kinetic energy of the chain with colliding kink and antikink as a function of time. In (a) $h=0.9$ and in (b) $h=1.3$, and in both cases the initial velocity of the kink and antikink is $v_{\rm c}^{}=0.9$. Collision takes place at $t=0$ and in both cases it results in formation of a DB.}
 \label{Fig:08}
\end{center}
\end{figure}

In Fig.~\ref{Fig:08} the kinetic energy of the chain is plotted as a function of time for the cases (a) $h=0.9$ and (b) $h=1.3$.
%%%%%%%%%%%%%%%%%%%%% Fig. 8
In both cases kink and antikink collide with the velocity $v_{\rm c}^{}=0.9$ at $t=0$ with the formation of DBs. Each collision of the kinks forming a DB results in a sharp double-peak of the kinetic energy. We estimated the oscillation period and then frequency of bions and DBs numerically by averaging over ten oscillations starting from the sixth one, see Fig.~\ref{Fig:08}. The first five oscillations are dropped because in some cases they have very long period, e.g., in multi-bounce collisions. For example, in Fig.~\ref{Fig:08}(a) ten oscillations take 53 time units, then $T=5.3$, and $\omega_{\rm DB}^{}=2\pi/T=1.18$, which is below the lower edge of the phonon band $\omega_1^{}=2$. The second harmonic has frequency $2\omega_{\rm DB}^{}=2.36$, which is above the upper edge of the phonon band $\omega_2^{}=2.22$. Note that the period of the bions in some cases varies in time noticeably [e.g., in the case presented in Fig.~\ref{Fig:05}(a)] due to the energy exchange between the kink's internal and translational modes and due to the losses to radiation. The period of the DBs is more robust since they radiate much less energy having no interaction with the phonons. 

\begin{figure}[t!]
\begin{center}
  \centering
          \subfigure[]{\includegraphics[width=0.49
 \textwidth,height=0.24\textheight]{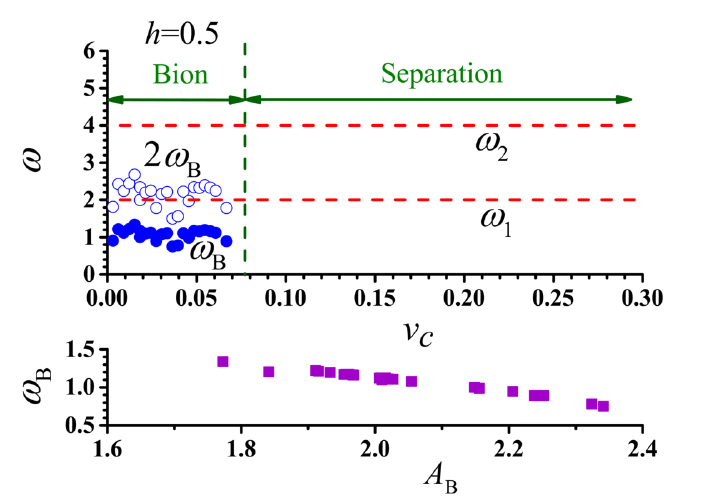}\label{fig:DBOmega_Amplitude_Vi_h_050}}
% \textheight]{fig9a.eps}\label{fig:DBOmega_Amplitude_Vi_h_050}}
%  \hspace{1mm}
          \subfigure[]{\includegraphics[width=0.49\textwidth,height=0.24\textheight]{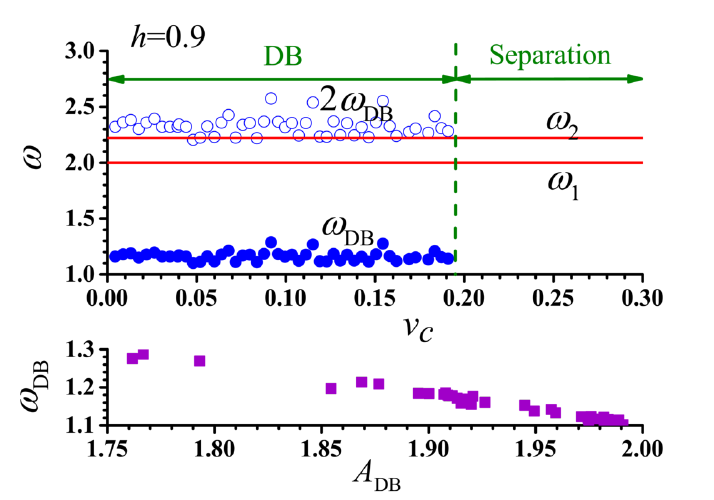}\label{fig:DBOmega_Amplitude_Vi_h_060}}
% \textheight]{fig9b.eps}\label{fig:DBOmega_Amplitude_Vi_h_060}}
%  \hspace{1mm}
        \subfigure[]{\includegraphics[width=0.49
 \textwidth,height=0.24\textheight]{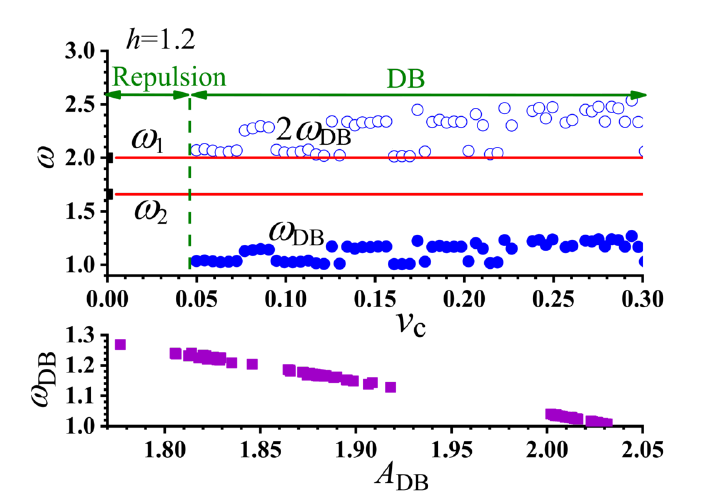}\label{fig:DBOmega_Amplitude_Vi_h_075}}
% \textheight]{fig9c.eps}\label{fig:DBOmega_Amplitude_Vi_h_075}}
%  \hspace{1mm}
      \subfigure[]{\includegraphics[width=0.49 \textwidth,height=0.24\textheight]{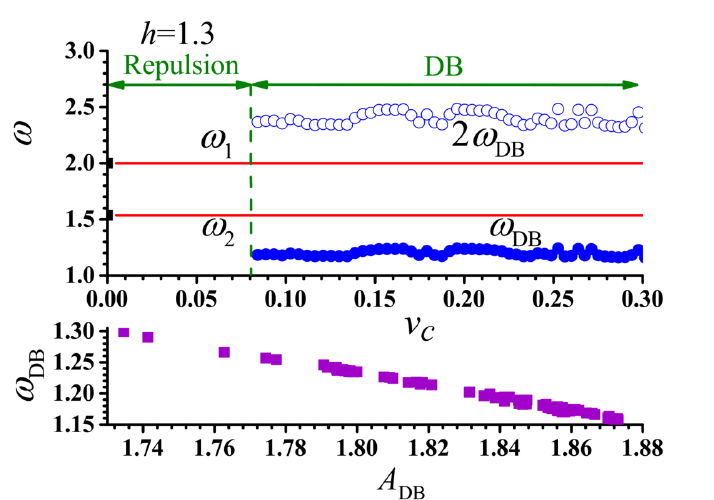}
 \label{fig:DBOmega_Amplitude_Vi_h_09}}
 \caption{Top panels: effect of the collision velocity $v_{\rm c}^{}$ on the collision outcome for (a) $h=0.5$, (b) $h=0.9$, (c) $h=1.2$, and (d) $h=1.3$. Horizontal dashed lines show the borders of the phonon spectrum, $\omega_1^{}$ and $\omega_2^{}$. Bion frequency $\omega_{\rm B}^{}$, or DB frequency $\omega_{\rm DB}^{}$, are shown by dots, and the second harmonics by circles. Bottom panels: (a) frequency of a bion formed in the collision with $v_{\rm c}^{}<0.07$ as a function of its amplitude; (b)--(d) frequency of DB, formed as a result of collision with sufficiently large velocity as a function of its amplitude.}
  \label{Fig:09}
\end{center}
\end{figure}

More information on the effect of the collision velocity on the collision outcome is presented in Fig.~\ref{Fig:09}.
%%%%%%%%%%%%%%%%%%%%% Fig. 9
For instance, for $h=0.5$ in the top panel of (a) the frequency of a bion formed when the collision velocity is relatively small is shown by dots as a function of $v_{\rm c}^{}$. The circles show the frequency of the bion's second harmonic. Horizontal dashed lines show the borders of the phonon spectrum frequencies, $\omega_1^{}$ and $\omega_2^{}$. It can be seen that the bion's frequency is always below the phonon spectrum but its second or third harmonic is within the spectrum. For $v_{\rm c}^{}>0.08=v^{*}$ the kink and antikink separate after the first collision. The bottom panel shows the bion frequency as a function of its amplitude for all the cases where a bion was formed for $h=1.2$ (left) and $h=1.3$ (right). Similar results are presented in (b) for $h=0.9$, but in this case not a bion but a DB is formed with the main frequency below the phonon band and all higher harmonics lie above the phonon band, hence suggesting the long time persistence of the structure.

In the particular case of $h=1$ the kink-antikink collisions result in the formation of DBs if the collision velocity is roughly below 0.25 and the kinks separate after the first collision for higher collision velocities, see Fig.~\ref{Fig:07}. Formation of DBs in the case of $h=1$ is not surprising because in this case, as it has been mentioned above, the width of the phonon band vanishes, hence DBs should generically survive.

\section{Conclusions and future work}\label{sec:conclusion}

Kink-antikink collisions were analyzed numerically in the discrete $\phi^4$ model, Eq.~\eqref{SpeightWardphi4}, which is free of the static Peierls--Nabarro potential. The lattice spacing $h$ of order unity was considered, which corresponds to a high discreteness with the kink spanning only just a few lattice sites, see Fig.~\ref{Fig:01}(a). Exact static kink solutions were derived iteratively from the two-point map \eqref{eq:phi4twopoint} starting from any admissible initial value $\phi_n^{}$. The existence of a one-parameter set of static kinks positioned arbitrarily with respect to the lattice ensures the existence of the zero frequency Goldstone translational mode, see Fig.~\ref{Fig:01}(c). The profile of this mode can be found by solving the eigenvalue problem for the equations of motion linearized near the static kink solution, see Eq.~\eqref{eq:Phi4Liniriazation}. The moving kink can be obtained by using the Goldstone mode as described in Sec.~\ref{sec:SetupBoosting}. Only symmetric collisions between the kink and its mirror image antikink moving with velocities $v_{\rm c}^{}$ and $-v_{\rm c}^{}$, respectively, were addressed. 

The borders of the phonon spectrum of the considered lattice, $\omega_1^{}$ and $\omega_2^{}$, cross at $h=1$, see Fig.~\ref{Fig:01}(b). Around $h=1$, the width of the relevant band is small (and it vanishes at $h=1$). The crossing of the phonon band edges changes the kink profile: for $h<1$ the kink's tails are monotonic but for $h>1$ they oscillate near the asymptotic values $\pm 1$, as follows from Eq.~\eqref{eq:tail}  and as can be seen in Fig.~\ref{Fig:01}(a). More importantly, for $h<1$ kink and antikink are mutually attractive topological solitons, while for $h>1$ they repel each other, as was shown in Sec.~\ref{sec:ForceCalc} and in our study of collisional dynamics in Sec.~\ref{sec:collisions}. The collision outcome is qualitatively different for attractive ($h<1$) and repulsive ($h>1$) kink and antikink, as summarized below.

For $h<1$ collisions with a velocity $v_{\rm c}^{}>v^*$ result in separation of the kink and antikink after the first collision, see Fig.~\ref{Fig:05}(a$^{\prime\prime}$) or (b$^{\prime\prime}$). The critical velocity $v^*$ decreases with increasing $h$ in the range $h<0.6$ and increases for larger $h$, see Fig.~\ref{Fig:07}. Collisions with a velocity smaller than $v^*$ can result in either separation of the kink and antikink after a multi-bounce collision, see Fig.~\ref{Fig:05}(a$^{\prime}$), or in the formation of an oscillatory mode. For $h<0.7$, when the phonon band is wide, the  oscillatory mode is a bion, see Fig.~\ref{Fig:05}(a), with frequency below the phonon band and higher harmonics within the band, see the upper panel of Fig.~\ref{Fig:09}(a). Due to the interaction with the phonons, the bion constantly radiates energy, its amplitude decreases and frequency increases, see lower panel of Fig.~\ref{Fig:09}(a). For $0.7<h<1$ the phonon band width is small and the oscillatory mode is a DB, see Fig.~\ref{Fig:05}(b) or (b$^{\prime}$), with the main frequency lying below the phonon band and all higher harmonics above the band as shown in the upper panel of Fig.~\ref{Fig:09}(b). The DB does not interact with the phonons and has a very long lifetime.

For $h>1$ the kink and antikink moving toward each other with a velocity $v_{\rm c}^{}<v^{**}$ cannot overcome the repulsion and they actually do not collide, see Fig.~\ref{Fig:05}(c), or (d), or (d$^{\prime}$). The critical velocity $v^{**}$ increases linearly with increasing $h$, see Fig.~\ref{Fig:07}. Collision with a velocity above $v^{**}$ can result in either separation of the kink and antikink after a single collision, see Fig.~\ref{Fig:05}(c$^{\prime\prime}$), or in formation of a DB, see Fig.~\ref{Fig:05}(c$^{\prime}$), or in a burst of radiation producing a rapidly decaying bion, see Fig.~\ref{Fig:05}(d$^{\prime\prime}$), where the radiation burst itself is not shown. A DB is formed for $1<h<1.5$ where the phonon band width is small. In this case the DB frequency is below the phonon band with all higher harmonics being above the band; see the upper panels of Fig.~\ref{Fig:09}(c) and (d). An energy burst is produced for $h>1.5$ when the colliding kink and antikink create an oscillatory mode with a frequency inside a relatively wide phonon band. Such a mode radiates energy very quickly and it has a very short lifetime.

Note that when the colliding kink and antikink produce an oscillatory mode, the mode has frequency within the range from 0.8 to 1.3 in a wide range of $h$, see Fig.~\ref{Fig:09}. In a forthcoming study properties of DBs will be addressed in detail regardless of the mechanism of their generation (generated not only in kink-antikink collisions) in the whole range of possible frequencies. So far DBs have not been analyzed in the discrete systems free of the static Peierls--Nabarro potential. One of the most intriguing features here is the DB mobility: it is relevant to understand whether it is enhanced or not due to the absence of the static Peierls--Nabarro potential. Contrary to more standard models, the present scenario has the potential of mobile breathers even for regimes of high discreteness, a feature that is uncommon for Klein--Gordon models outside the realm of integrable systems.

Another direction for future studies is the analysis of kink collisions in the $\phi^6$ and $\phi^8$ models \cite{Gani.PRD.2014,Gani.JHEP.2015,Moradi.JHEP.2017,Belendryasova2019,Demirkaya2017,Gani.arXiv.2020.explicit}. Interestingly, kinks in the $\phi^6$ model are asymmetric and have short-range tails \cite{Gani.PRD.2014,Moradi.JHEP.2017,Demirkaya2017}. In the $\phi^8$ and a number of suitable higher-order models the kinks can have long-range tails with power-law decay \cite{Belendryasova2019,Christov.PRD.2019,Christov.PRL.2019,Manton2019,Khare.JPA.2019}. It would be interesting to study kink collisions in these models in the regime of high discreteness. In the work \cite{Rakhmatullina} two discrete $\phi^6$ models free of the static Peierls--Nabarro potential have been derived. Generalizing such a derivation to $\phi^{8}$, $\phi^{10}$ and $\phi^{12}$ would enable the consideration of the intriguing interplay of discreteness and long-range interactions. A discrete realm may be more straightforward of a place to consider such long-range interactions given that extended lattices would be easier to consider than the considerably more computationally expensive continuum analogues thereof.

The third natural continuation of this work is the analysis of the multi-bounce collisions and the related analysis of the kink's internal modes. As it can be seen from Fig.~\ref{Fig:01}(b), in the case of small $h$ there is only one kink's internal mode with the frequency $\omega \approx \sqrt{3}$, but for $h\sim 1$ the kink has more than one internal modes with frequencies below the phonon spectrum. This fact should affect the picture of multi-bounce collisions and this was indeed observed in our preliminary simulations. 

\section*{Acknowledgments}

The work of the MEPhI group was supported by the MEPhI Academic Excellence Project (Contract No.\ 02.a03.21.0005, 27.08.2013). V.A.G.\ and S.V.D.\ acknowledge the support of the Russian Foundation for Basic Research, Grant No.\ 19-02-00971. The work was partly supported by the State assignment of IMSP RAS. PGK gratefully acknowledges the hospitality of the Mathematical Institute of the University of Oxford and the support of the Leverhulme Trust during the final stages of this work. This material is based upon work supported by the US National Science Foundation under Grant DMS-1809074 (PGK).

\end{document}